\documentclass[twocolumn,preprintnumbers,nofootinbib,amsmath,amssymb,floatfix]{revtex4}

\usepackage{graphicx}
\usepackage{dcolumn}
\usepackage{bm}
\usepackage{ifpdf}
\newcommand {\vtwoobspaper}{\vtwo^{\text{obs}}}
\newcommand {\meanvtwoflow}{\mean{\vtwo}_{\text{flow}}}

\ifpdf
\usepackage[linkbordercolor={0 0 .8},%
             citebordercolor={.8 0 0},%
             urlbordercolor={.8 0 .8},%
             raiselinks=true,%
             pdfborder={0 0 1 [3]}]{hyperref}
\pdfoutput=1        
\pdfcatalog{/PageMode(/UseOutlines)} 
\fi

\def\phob         {PHOBOS}
\def\prap         {pseudorapidity}

\newcommand {\vtwoetaf}[1]{\vtwo(\eta_{#1})^{\text{fit}}}

\newcommand {\hij}{HIJING}

\newcommand {\pyth}{PYTHIA}

\newcommand {\vtwo}{v_2}
\newcommand {\vtwosqr}{v_2^2(\eta_1,\eta_2)}

\newcommand {\sigmatot}{\sigma_{\text{dyn}}}

\newcommand {\sigmaflow}{\sigma_{\text{flow}}}
\newcommand {\sigmadelta}{\sigma_\delta}
\newcommand {\sigmastat}{\sigma_n}

\newcommand {\mean}[1]{\left\langle #1 \right\rangle}

\newcommand {\der}{{\rm{d}}}
\newcommand {\BesselGaus}[3]{{\rm BG}\left(#1;#2,#3\right)}

\newcommand {\detagt}{|\deta| \! > \! 2}
\newcommand {\detalt}{|\deta| \! < \! 2}

\newcommand {\expt}{\text{exp}}

\def\etal         {et al.}

\newcommand {\snn}      {\sqrt{s_{\scriptscriptstyle{{\rm NN}}}}}

\newcommand {\abs}[1]   {\ensuremath{\left| #1 \right|}}

\newcommand {\fig}[1]{Fig.~\ref{#1}}

\newcommand {\eq}[1]{Eq.~\ref{#1}}
\newcommand {\eqs}[2]{Eq.~\ref{#1} and~\ref{#2}}
\newcommand {\Eq}[1]{Equation~\ref{#1}}

\newcommand {\sect}[1]{Sect.~\ref{#1}}

\newcommand {\Sect}[1]{Section~\ref{#1}}
\newcommand {\Sects}[2]{Sections~\ref{#1} and~\ref{#2}}

\newcommand {\pt}       {\ensuremath{p_{\mathrm{T}}}}

\newcommand {\Npart}    {\ensuremath{N_{\rm part}}}

\newcommand {\deta}		{\ensuremath{\Delta\eta}}
\newcommand {\dphi}		{\ensuremath{\Delta\phi}}

\newcommand {\gev}   {\mbox{${\rm GeV}$}}

\newcommand {\mom}   {\mbox{\rm GeV$\kern-0.15em /\kern-0.12em c$}}
\newcommand {\mmom}  {\mbox{\rm MeV$\kern-0.15em /\kern-0.12em c$}}
\newcommand {\mass}  {\mbox{\rm GeV$\kern-0.15em /\kern-0.12em c^2$}}
\newcommand {\mmass} {\mbox{\rm MeV$\kern-0.15em /\kern-0.12em c^2$}}

\newcommand {\cm}    {\mbox{${\rm cm}$}}

\newcommand {\AuAu}  {\mbox{Au+Au}}

\begin{document}

\title{Non-flow correlations and elliptic flow fluctuations \\in Au+Au collisions at 
\mbox{\boldmath$\snn$}=~200~GeV}
\author{
%
%
B.Alver$^4$,
B.B.Back$^1$,
M.D.Baker$^2$,
M.Ballintijn$^4$,
D.S.Barton$^2$,
R.R.Betts$^6$,
A.A.Bickley$^7$,
R.Bindel$^7$,
W.Busza$^4$,
A.Carroll$^2$,
Z.Chai$^2$,
M.P.Decowski$^4$,
E.Garc\'{\i}a$^6$,
T.Gburek$^3$,
N.George$^2$,
K.Gulbrandsen$^4$,
C.Halliwell$^6$,
J.Hamblen$^8$,
M.Hauer$^2$,
C.Henderson$^4$,
D.J.Hofman$^6$,
R.S.Hollis$^6$,
R.Ho\l y\'{n}ski$^3$,
B.Holzman$^2$,
A.Iordanova$^6$,
E.Johnson$^8$,
J.L.Kane$^4$,
N.Khan$^8$,
P.Kulinich$^4$,
C.M.Kuo$^5$,
W.Li$^4$,
W.T.Lin$^5$,
C.Loizides$^4$,
S.Manly$^8$,
A.C.Mignerey$^7$,
R.Nouicer$^{2,6}$,
A.Olszewski$^3$,
R.Pak$^2$,
C.Reed$^4$,
C.Roland$^4$,
G.Roland$^4$,
J.Sagerer$^6$,
H.Seals$^2$,
I.Sedykh$^2$,
C.E.Smith$^6$,
M.A.Stankiewicz$^2$,
P.Steinberg$^2$,
G.S.F.Stephans$^4$,
A.Sukhanov$^2$,
M.B.Tonjes$^7$,
A.Trzupek$^3$,
C.Vale$^4$,
G.J.van~Nieuwenhuizen$^4$,
S.S.Vaurynovich$^4$,
R.Verdier$^4$,
G.I.Veres$^4$,
P.Walters$^8$,
E.Wenger$^4$,
F.L.H.Wolfs$^8$,
B.Wosiek$^3$,
K.Wo\'{z}niak$^3$,
B.Wys\l ouch$^4$\\
\vspace{3mm}
\small
%
%
 $^1$~Physics Division, Argonne National Laboratory, Argonne, IL 60439-4843,
 USA\\
 $^2$~Physics and C-A Departments, Brookhaven National Laboratory, Upton, NY
 11973-5000, USA\\
 $^3$~Institute of Nuclear Physics PAN, Krak\'{o}w, Poland\\
 $^4$~Laboratory for Nuclear Science, Massachusetts Institute of Technology,
 Cambridge, MA 02139-4307, USA\\
 $^5$~Department of Physics, National Central University, Chung-Li, Taiwan\\
 $^6$~Department of Physics, University of Illinois at Chicago, Chicago, IL
 60607-7059, USA\\
 $^7$~Department of Chemistry, University of Maryland, College Park, MD 20742,
 USA\\
 $^8$~Department of Physics and Astronomy, University of Rochester, Rochester,
 NY 14627, USA\\
}

\begin{abstract}\noindent
  This paper presents results on event-by-event elliptic flow
  fluctuations in Au+Au collisions at $\snn =$
  200~GeV, where the contribution from non-flow correlations has been
  subtracted.  An analysis method is introduced to measure non-flow
  correlations, relying on the assumption that non-flow correlations
  are most prominent at short ranges ($\detalt$). Assuming that
  non-flow correlations are of the order that is observed in p+p
  collisions for long range correlations ($\detagt$), relative
  elliptic flow fluctuations of approximately 30-40\% are observed.
  These results are consistent with predictions based on spatial
  fluctuations of the participating nucleons in the initial nuclear
  overlap region. It is found that the long range non-flow
  correlations in Au+Au collisions would have to be more than an order of
  magnitude stronger compared to the p+p data to lead to the observed
  azimuthal anisotropy fluctuations with no intrinsic elliptic flow
  fluctuations.
\vspace{3mm}
\noindent 
\end{abstract}

\maketitle

\section{Introduction}
\label{intro}

The characterization of the collective flow of produced particles by
their azimuthal anisotropy has proven to be one of the more fruitful
probes of the dynamics of heavy ion collisions at the Relativistic
Heavy Ion Collider (RHIC). Flow is sensitive to the early stages of
the collision and so the study of flow affords unique insights into
the properties of the hot and dense matter that is produced, including
information about the degree of thermalization and its equation of
state~\cite{Kolb:2000fha}.

Elliptic flow, quantified by the second coefficient, $v_2$, of a
Fourier decomposition of the azimuthal distribution of observed
particles relative to the event-plane angle, has been studied
extensively in collisions at RHIC as a function of \prap,
centrality, transverse momentum, center-of-mass energy and system
size~\cite{Back:2004zg, Back:2004mh, Alver:2006wh, Back:2004je, Adams:2005dq,
  Adcox:2004mh}. A detailed comparison of these results to theoretical
models requires a quantitative understanding of the contributions of
other many-particle correlations, referred to as ``non-flow'' and
event-by-event elliptic flow fluctuations~\cite{Song:2008hj}. In
particular, the measurement of event-by-event fluctuations can pose
new constraints on the models of the initial state of the collision
and its subsequent hydrodynamic 
evolution~\cite{Osada:2001hw, Alver:2008zza}.

Comparison of the elliptic flow measurements in the Au+Au and Cu+Cu
systems at RHIC suggests the existence of large fluctuations in the
initial geometry of heavy ion collisions~\cite{Alver:2006wh}. These
initial state fluctuations are expected to lead to event-by-event
fluctuations in the measured elliptic flow signal. The measurement in
Au+Au collisions of dynamic fluctuations in $v_2$, including
contributions from event-by-event elliptic flow fluctuations and
non-flow correlations, has yielded results which are consistent with
this expectation~\cite{Alver:2007qw}.

Different methods have been proposed to reduce the contribution of
non-flow correlations to the elliptic flow
measurements~\cite{Borghini:2001vi,Bilandzic:2008nx}.  However, the
application of these methods to the measurement of elliptic flow
fluctuations is limited due to the complicated interplay between non-flow correlations and elliptic flow fluctuations~\cite{Borghini:2001vi,Alver:2008zza}.  

Ollitrault \etal\ have suggested estimating the magnitude of non-flow
from measurements of correlations in p+p
collisions~\cite{Ollitrault:2009ie}.  However, this estimation may not
be completely reliable since a richer correlation structure is
observed in Au+Au collisions at RHIC in comparison to the p+p
system~(e.g.~\cite{Adams:2004pa,Alver:2008gk, Abelev:2009qa,
  Alver:2009id}). We propose a method to
separate flow and non-flow contributions to the second Fourier
coefficient of azimuthal particle pair distributions by studying the
three-dimensional two-particle correlation function in
($\eta_1,\eta_2,\dphi$) space. This separation relies on the
assumption that non-flow correlations are most prominent in short
range (\mbox{$\deta \! \equiv \!  |\eta_1-\eta_2| \! < \!  2$}). 
The presumably small long
range ($\detagt$) non-flow correlations are estimated using p+p data,
and \hij\ and \pyth\ models. 
Estimation of non-flow correlations using these assumptions
allows the subtraction of the
contribution of non-flow correlations to the measured dynamic $v_2$
fluctuations to obtain event-by-event elliptic flow fluctuations.

This paper is organized as follows. The experimental data is described
in \Sect{sect:dataset}. The measurement of the non-flow correlations
and the corresponding event-by-event elliptic flow fluctuations are
presented in \Sects{sect:nonflow}{sect:flow}. Discussion and
conclusions are included in \sect{sect:conclude}. The numerical
relation between dynamic $v_2$ fluctuations, elliptic flow
fluctuations and non-flow correlations is addressed in
Appendix~\ref{app:relate}.

\section{Experimental data}
\label{sect:dataset}

The data presented here for Au+Au collisions at $\snn =$ 200~GeV were
collected during RHIC Run 4 (2004) using the \phob\
detector~\cite{Back:2003sr}. The primary event trigger requires a
coincidence between the Paddle Counters, which are two sets of sixteen
scintillator detectors located at $3.2 < |\eta| < 4.5$.  
An online vertex is determined from the time difference between
signals in two sets of 10 Cerenkov counters located at 4.4 $<|\eta|<$
4.9, to select collisions that are close to the nominal vertex
position $z_{vtx}=0$ along the beam-axis.

Offline vertex reconstruction makes use of information from different
sub-detectors. Two sets of double-layered silicon Vertex Detectors
(VTX) are located below and above the collision point. \phob\ also has
two Spectrometer arms in the horizontal plane used for tracking and
momentum measurement of charged particles. For events in the selected
vertex region, the most accurate $z$ (along the beam) 
and $y$ (vertical, perpendicular to the beam) positions are
obtained from the Vertex Detector, while the position along $x$
(horizontal, perpendicular to the beam) comes primarily from the
Spectrometer.  

The collision centrality is defined through bins of fractional total
inelastic cross section, determined using the energy deposited in the
Paddle Counters. In this paper, we report results for 6--45\% most
central events, for which measured dynamic $v_2$ fluctuations
values are available~\cite{Alver:2007qw}. About 4 million collision
events were selected in this centrality range by requiring that the
primary collision vertex falls within $|z_{vtx}|<$ 6~cm.

The analysis presented in this paper is performed using the
reconstructed hits in the large-acceptance \phob\ Octagon silicon
array, covering \prap\ \mbox{$-3\!<\!\eta<3$} over almost the full
azimuth. The angular coordinates ($\eta,\phi$) of charged particles
are measured using the location of the energy deposited in the
single-layer silicon pads of the Octagon.  After merging of signals in
neighboring pads, in cases where a particle travels through more than
a single pad, the deposited energy is corrected for the angle of
incidence, assuming that the charged particle originated from the
primary vertex.  Noise and background hits are rejected by placing a
lower threshold on the corrected deposited energy. Depending on
$\eta$, merged hits with less than 50-60\% of the energy loss expected
for a minimum ionizing particle are rejected~\cite{Back:2001bq}.
Since the multiplicity array consists of single-layer silicon
detectors, there is no $p_{T}$, charge or mass information available
for the particles. All charged particles above a low-$p_{T}$ cutoff of
about 7~MeV/c at $\eta$=3, and 35~MeV/c at $\eta$=0 (which is the
threshold below which a charged pion is stopped by the beryllium
beam pipe) are included on equal footing.

\section{Measurement of non-flow correlations} 
\label{sect:nonflow}
If the only correlations between particles are due to elliptic flow,
then the distribution of the azimuthal angular separation between
particles ($\dphi \! \equiv \!  \phi_1-\phi_2$) is given by
$1+2V\cos(2\dphi)$, where $V=\vtwo(\eta_1)\times \vtwo(\eta_2)$. In
general, the second Fourier coefficient of the $\dphi$ distribution
has contributions from both flow and non-flow correlations. 

Flow and non-flow contributions can be separated with a detailed study
of the $\eta$ and $\deta$ dependence of the $\dphi$ correlation
function. Consider the distribution of $\dphi$ between particles
selected from two $\eta$ windows centered at $\eta_1$ and $\eta_2$. 
We define the
quantity $\vtwosqr$ as the sum of flow and non-flow contributions to
the second Fourier coefficient of the normalized $\dphi$ distribution:
\begin{equation}
  \vtwosqr \equiv \mean{\cos(2\dphi)}(\eta_1,\eta_2)
\end{equation} 

The contributions to the second Fourier coefficient of the $\dphi$
distribution can be parameterized as
\begin{equation}
  \mean{\cos(2\dphi)} = \mean{\vtwo^2}_{\text{flow}} + \delta,
\end{equation}
where $\delta$ is the contribution of non-flow correlations~\cite{Poskanzer:1998yz}.
Using the fact that elliptic flow leads to a correlation between all
particles in the event and creates a signal which only depends on
\prap\ ($\vtwo(\eta)$), we can write:
\begin{equation}
  \vtwosqr = \vtwo(\eta_1)\!\times\!\vtwo(\eta_2) + \delta(\eta_1,\eta_2),
  \label{eq:v2sqr}
\end{equation}
The measurement of non-flow correlations is therefore achieved in two
steps, described in the following sections. 
First we measure the three dimensional ($\eta_1, \eta_2,
\dphi$) correlation function to obtain $\vtwosqr$. Then we separate
the observed $\vtwosqr$ distribution to its flow and non-flow
components.

\subsection{Two particle correlations analysis}
Two particle correlations have been studied extensively in ($\deta,
\dphi$) space using the \phob\ detector for various collision
systems~\cite{Alver:2007wy,Alver:2008gk}. In this analysis, we extend
the same analysis procedure to ($\eta_1,\eta_2,\dphi$) space.

The inclusive two-particle correlation function in
($\eta_1,\eta_2,\dphi$) space is defined as follows
\begin{equation}
\label{2pcorr_incl}
R_n(\eta_1,\eta_2,\dphi)=\mean{\frac{\rho_{n}^{\rm II}(\eta_1,\eta_2,\dphi)}
  {\rho^{\rm mixed}(\eta_1,\eta_2,\dphi)}-1}
\end{equation} 
where $\rho_{n}^{\rm II}(\eta_1,\eta_2,\dphi)$ (with unit integral in
each $\eta_1,\eta_2$ bin) is the foreground pair distribution obtained
by taking two particles from the same event, then averaging over
all pairs in all events and $\rho^{\rm mixed}(\eta_1,\eta_2,\dphi)$ (with unit
integral in each $\eta_1,\eta_2$ bin) is the mixed-event background
distribution constructed by randomly selecting two particles from two
different events with similar vertex position and centrality,
representing a product of two single particle distributions.  A vertex
bin size of 0.2~cm is used in the event-mixing.
 
The high occupancies measured in A+A collisions require us to account
for the high probability of multiple particles hitting a single pad.
Furthermore, secondary effects, such as $\delta$-electrons, $\gamma$
conversions and weak decays, cannot be all rejected directly.
Corrections for the high occupancy in the Octagon detector and the
secondary effects have been applied in the same way as in the
previous $\deta,\dphi$ correlation
analyses~\cite{Alver:2007wy,Alver:2008gk}.

To correct for the effects of occupancy, each hit is assigned a weight
while calculating the correlation function. The weight is calculated
using the centrality of the event and \prap\ of the hit (which
determine the likelihood of multiple particles passing through a pad for a given
$\der E/\der x$ value) and the $\der E/\der x$ information. The
details of the occupancy correction can be found in
Ref.~\cite{Alver:2008gk}.

To correct for the secondary detector effects in the data, correlation
functions were calculated for different Monte Carlo event generators
(\pyth, \hij\ and a modified \pyth\ in which all intrinsic
correlations have been removed) at $\snn$ = 200~GeV both at
the generator level for true primary charged hadrons and with the full
GEANT detector simulation and reconstruction procedure. The overall
correlation structure for the reconstructed Monte Carlo events
consists of both intrinsic and secondary correlations and these two
sources of correlations were found to be largely independent of each
other, i.e.\ the correlation from secondaries is mostly determined by
sensor thickness, detector geometry, known cross-sections and decay
kinematics~\cite{Alver:2007wy}. 

The final correlation function,
$R_{n\text{ final}}^{\rm data}(\eta_1,\eta_2,\dphi)$ is calculated
from the raw correlation function, $R_{n\text{ raw}}^{\rm
  data}(\eta_1,\eta_2,\dphi)$ by subtracting the contribution from
secondary correlations:
\begin{multline}
R_{n\text{ final}}^{\rm data}(\eta_1,\eta_2,\dphi) = \\
R_{n\text{ raw}}^{\rm data}(\eta_1,\eta_2,\dphi)
-S(\eta_1,\eta_2,\dphi),
\label{eq:Rcorrection}
\end{multline}
where the correction factor $S(\eta_1,\eta_2,\dphi)$ is calculated by
comparing the generator level correlation function excluding
particles outside the \phob\ detector acceptance, $R_{n \text{
    pri,acc}}^{\rm MC}(\eta_1, \eta_2, \Delta \phi)$, to the
correlation function obtained with the full GEANT detector simulation
and reconstruction procedure, $R_{n \text{ sim}}^{\rm MC}(\eta_1,
\eta_2, \Delta \phi)$:
\begin{multline}
  \label{eq:Scorrection}
  S(\eta_1,\eta_2,\dphi) = \\
  R_{n \text{ sim}}^{\rm MC}(\eta_1,\eta_2, \dphi)
  -R_{n \text{ pri,acc}}^{\rm MC}(\eta_1, \eta_2, \dphi).
\end{multline}
The correction factor $S(\eta_1,\eta_2,\dphi)$ is calculated
separately for each centrality bin using a set of \hij\ events with
appropriate average multiplicity. 
More details on the correction factor $S$ and its dependence on
$\deta$ and $\dphi$ can be found in~\cite{Alver:2007wy}.

\begin{figure}[t]
  \includegraphics[width=0.47\textwidth]{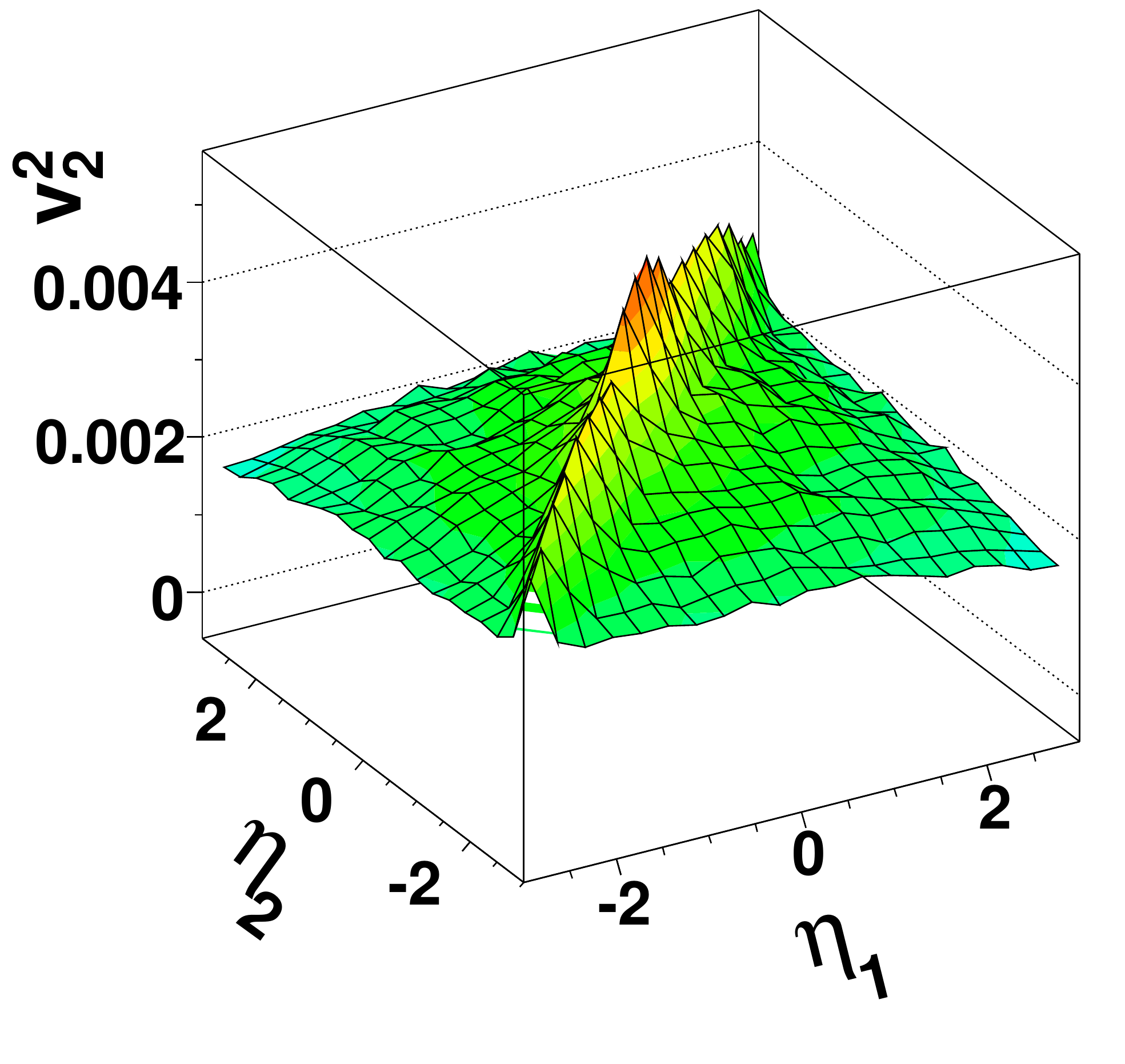}
  \caption{Second Fourier coefficient of the correlation function
    $R_n(\Delta\phi,\eta_1,\eta_2)$ as a function of $\eta_1$ and $\eta_2$
    for the 40-45\% central Au+Au collisions at $\snn
    =$~200~GeV. The ridge along $\eta_1=\eta_2$ represents the region
    where non-flow contributions are most prominent.}
  \label{fig:datasamplev2sqr}
\end{figure}

The second Fourier coefficient of the normalized $\dphi$ distribution
is calculated from the correlation function by a fit in each
($\eta_1,\eta_2$) bin:
\begin{equation}
  R_{n\text{ final}}^{\rm data}(\eta_1,\eta_2,\dphi) = 2 \vtwosqr \cos(2\dphi).
\end{equation}
The value of $\vtwosqr$ can also be calculated directly as
\begin{equation}
 \vtwosqr = \int R_{n\text{ final}}^{\rm data}(\eta_1,\eta_2,\dphi) \cos(2\dphi) \der \dphi.
\end{equation}
The two methods of calculating $\vtwosqr$ are found to be equivalent within the systematic uncertainties of the measurement.
The resulting $\vtwosqr$ distribution for 40-45\% centrality bin
is shown in \fig{fig:datasamplev2sqr}.

\begin{figure*}[t]
  \centering 
  \includegraphics[width=0.94\textwidth]{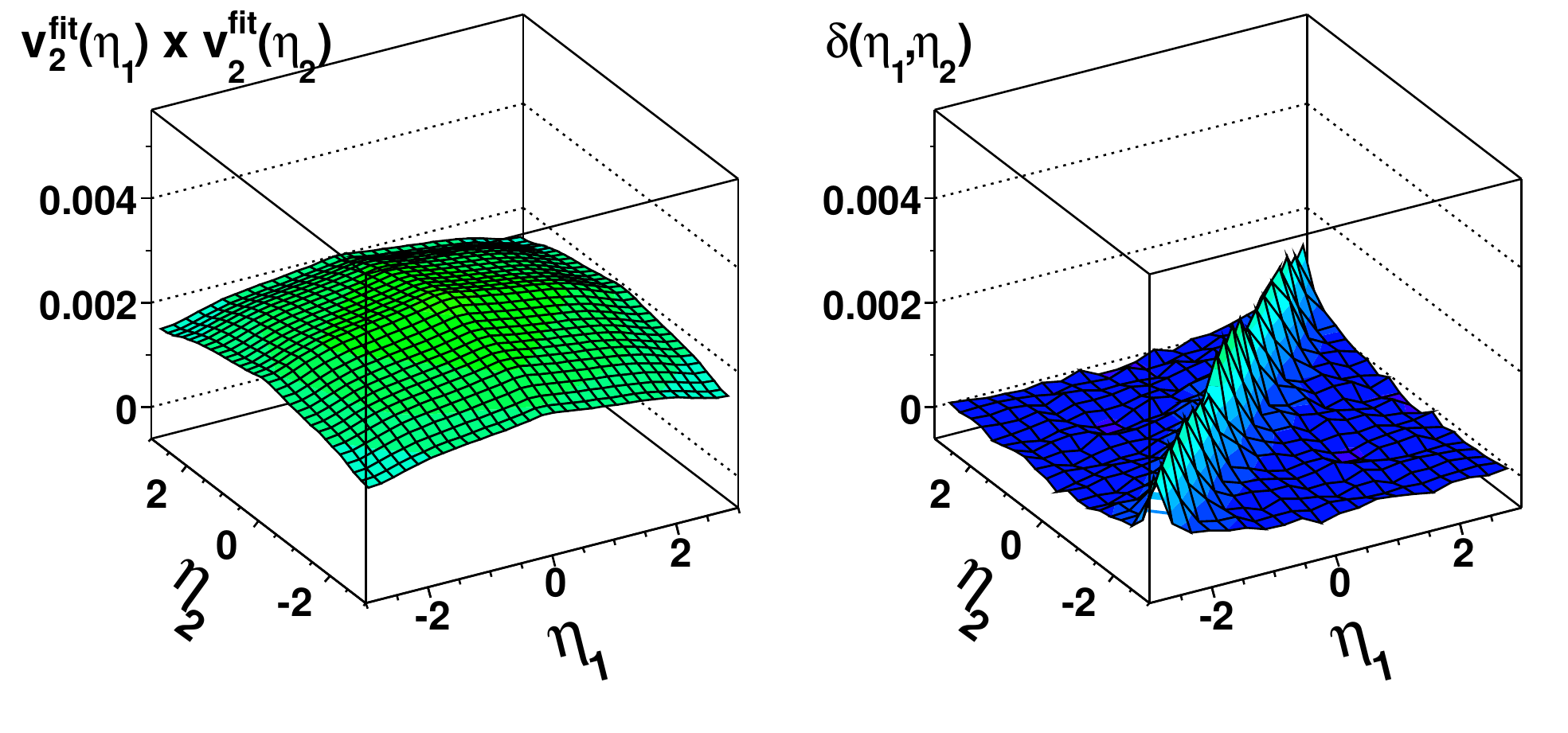}
  \caption{Flow (left) and non-flow (right) components of $\vtwosqr$ in \fig{fig:datasamplev2sqr} obtained by \eqs{eq:v2sqrfit}{eq:subtracttogetdelta} assuming non-flow correlations at $\detagt$ are negligible.}
  \label{fig:datasamplev2sqrfit}
\end{figure*}

\subsection{Separation of flow and non-flow contributions}
The measured $\vtwosqr$ signal in \fig{fig:datasamplev2sqr} shows the
features expected from \eq{eq:v2sqr}: a ridge along \mbox{$\deta=0$}
where the non-flow signal is most prominent which sits on a plateau
which can be factorized in $\eta_1$ and $\eta_2$. Assuming non-flow
correlations are small at large $\deta$ separations, it is possible to
separate the $\vtwosqr$ to its flow and non-flow components.

We start by assuming that non-flow correlations at $\detagt$
($\delta_{\detagt}$) are zero. Then, we can perform a fit
\begin{equation}
  \vtwosqr = \vtwoetaf{1}\times \vtwoetaf{2} \: ; \quad \abs{\eta_1-\eta_2}>2
\label{eq:v2sqrfit},
\end{equation}
where the fit function $\vtwoetaf{}$ is an eighth order even polynomial.  
The fit
in the selected $\deta$ region can be used to extract the magnitude of
correlations due to flow, $\vtwoetaf{1}\times \vtwoetaf{2}$, in the
whole \prap\ acceptance. Subtracting the correlations due to flow, we
can extract the contribution of non-flow correlations:
\begin{equation}
  \delta(\eta_1,\eta_2)=\vtwosqr-\vtwoetaf{1}\times \vtwoetaf{2}.
\label{eq:subtracttogetdelta}
\end{equation}
The two components of the $\vtwosqr$ distribution in
\fig{fig:datasamplev2sqr} are shown in \fig{fig:datasamplev2sqrfit}.

Different flow measurements with different methods and \prap\
acceptances are influenced differently by the non-flow correlation
signal. To calculate the effects of non-flow correlation on the
measurement of dynamic $v_2$ fluctuations
performed by \phob~\cite{Alver:2007qw}, we
calculate the average of the $\delta(\eta_1,\eta_2)$ and $\vtwosqr$
distributions over all particle pairs:
\begin{eqnarray} 
\mean{\delta} &=& 
  \frac{\int\delta(\eta_1,\eta_2)\frac{\der N}{\der \eta_1}\frac{\der N}{\der \eta_2} \der \eta_1 \der \eta_2}
  {\int\frac{\der N}{\der \eta_1} \frac{\der N}{\der \eta_2} \der \eta_1 \der \eta_2}   \label{eq:averagedelta}\\
\mean{v_2^2} &=& 
\frac{\int\vtwosqr\frac{\der N}{\der \eta_1}\frac{\der N}{\der \eta_2} \der \eta_1 \der \eta_2}
  {\int\frac{\der N}{\der \eta_1} \frac{\der N}{\der \eta_2} \der \eta_1 \der \eta_2},
  \label{eq:averagev22}
\end{eqnarray} 
where $\der N /\der \eta$ is the observed charged-particle
\prap\ distribution in the \phob\ detector. To cancel scale
uncertainties in these quantities, we calculate the ``non-flow ratio''
given by $\mean{\delta}/\mean{v_2^2}$.

\begin{figure*}[t]
  \centering \includegraphics[width=0.94\textwidth]{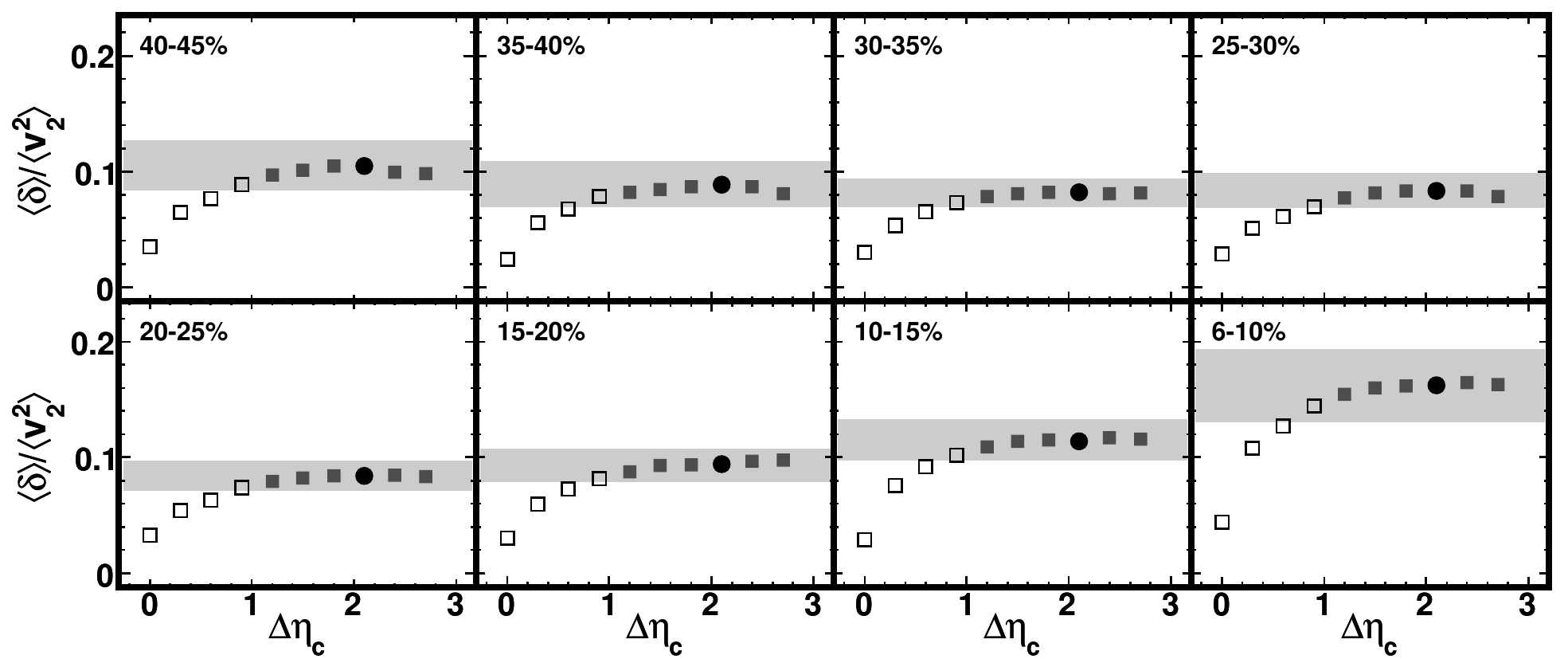}
  \caption{Measured value of the non-flow ratio
    ($\mean{\delta}/\mean{v_2^2}$) as a function of the $\deta$ cut
    ($\deta_c$) where non-flow correlations are assumed to be zero for
    $\abs{\deta}>\deta_c$ for different centrality bins. The black circles (one for each panel)
    show values for $\deta_c=2.1$ with the gray band denoting the 90\%
    C.L.\ systematic errors on those results as described in the text.
    The gray squares show values for $1.2\leq\deta_c\leq2.7$, which
    are used in the systematic error estimation. Open squares show
    values for $\deta_c<1$.}
  \label{fig:deltavsdeletadata}
\end{figure*}

The systematic uncertainty has been evaluated for the various stages
of the non-flow ratio calculation including the calculation of the
correlation function and the fit to $\vtwosqr$ to obtain the non-flow
ratio. A ``digital'' occupancy correction with only the event-by-event
hit density distribution and no $\der E/\der x$ information has been
used. Hits on the \phob\ Vertex detector, which has a different
granularity from the Octagon detector have been added to the
analysis. Monte Carlo samples with different average multiplicity from
the data have been used in the correction procedure. The $\deta$ cut
used in the fit has been varied between $1.2$ and $2.7$~\footnote{ The
  Octagon detector with a \prap\ coverage of \mbox{$-3\!<\!\eta<3$}
  allows particle pairs to be studied up to $\deta=6$. However, in
  this study the $\deta$ cut is constrained to $\deta_c<3$ such that
  particles from all $\eta$ values contribute in the fit to obtain
  $v_2(\eta)$.}.  Different fit functions $\vtwoetaf{}$ have been used
from second order up to eighth order polynomials. Finally the complete
analysis chain has been performed by dividing the data set into
$6\times2\cm$ wide vertex bins. Systematic errors are estimated for
different steps in the analysis using the variation in the results with
respect to the baseline due to these changes in the analysis.  The
errors in the different steps are added in quadrature to obtain the
90\% confidence interval on the measurement of non-flow ratio.

So far, we have assumed that long range ($\detagt$) non-flow
correlations can be neglected. However, studies of the correlation
function in p+p collisions show that non-flow correlations do extend
out to $\detagt$ in elementary
collisions~\cite{Alver:2007wy}. Furthermore, a rich correlation
structure in high \pt-triggered correlations that extend out to
$\detagt$ has been observed in 200~\gev\ \AuAu\ collision at
RHIC~\cite{Alver:2009id} after the estimated flow signal is
subtracted. However, 
due to the inherent uncertainty in the flow
subtraction, it is not possible to determine the second Fourier
coefficient of this correlation structure precisely.

The study of the non-flow ratio as a function of the $\deta$ cut
($\deta_c$) for the $\vtwoetaf{}$ fit carries important information on
the magnitude of non-flow at large $\deta$ separations. If non-flow
correlations are short ranged, we expect that the fits should yield
non-flow ratio results that saturate for large values of
$\deta_c$. The extracted value of $\mean{\delta}/\mean{v_2^2}$ is
plotted as a function of the $\deta_c$, where it is assumed that
$\delta$ is zero for $\abs{\deta}>\deta_c$, for different centrality
bins in \fig{fig:deltavsdeletadata}. The saturation expected if
non-flow correlations are short-range is indeed observed.  However, it
should be noted that the same saturation pattern could also be
observed with a finite magnitude of non-flow that has little $\deta$
dependence in the region $\deta>1.2$.

\begin{figure}[t]
  \includegraphics[width=0.47\textwidth]{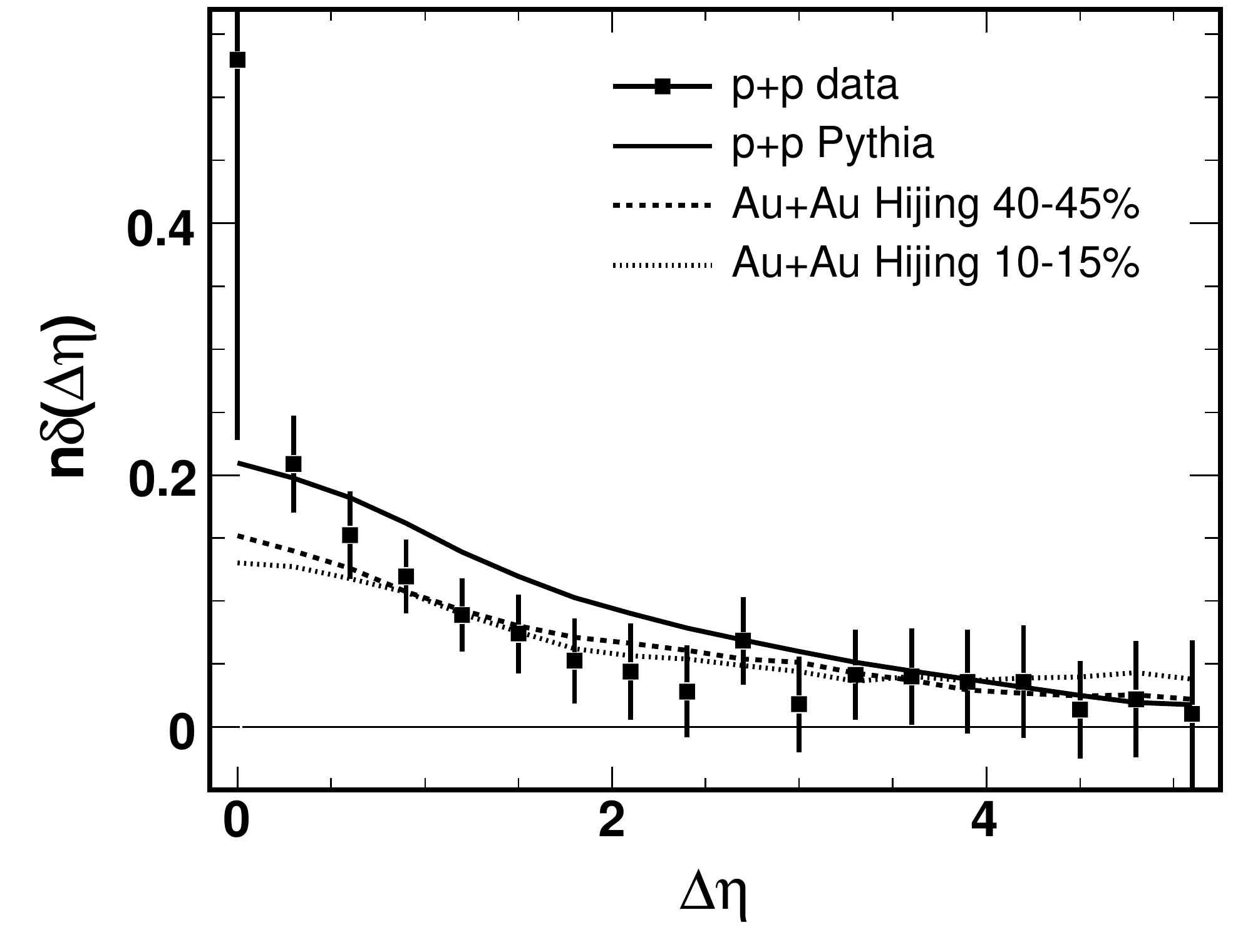}
  \caption{The magnitude of non-flow correlations ($\delta$) scaled by
    the charged particle multiplicity ($n$) in the \prap\ range
    $\abs{\eta}<3$ as a function of particle pair \prap\ separations
    ($\deta$) for p+p data and different Monte Carlo generators with
    no flow correlations at $\snn =$~200~GeV. The results for p+p data
    (squares) with 90\% C.L.\ systematic errors are obtained from two
    particle $\deta,\dphi$
    correlations~\cite{Alver:2007wy}. Statistical errors are not
    shown.}
  \label{fig:deltavsdeleta}
\end{figure}

To quantitatively assess the effect of non-zero non-flow correlations
at large $\deta$ separations, we analyze the correlation functions
obtained from Monte Carlo event generators. In p+p collisions, the
magnitude of non-flow correlations, $\delta$, can be directly
calculated as the second Fourier coefficient of $\dphi$ correlations
since elliptic flow is not present~\cite{Alver:2007wy}. If A+A
collisions were a superposition of p+p collisions, the value of
$\delta$ would be diluted due to the presence of uncorrelated
particles. To compare the strength of non-flow correlations in \hij\
(Au+Au) and \pyth\ (p+p) models and p+p collisions, we calculate the
value of $\delta$ scaled by the average event multiplicity, shown in
\fig{fig:deltavsdeleta}\footnote{The large uncertainty in the p+p data
  at $\deta=0$ is due to $\delta$-electrons and $\gamma$ conversions,
  which may not be completely described by GEANT
  simulations~\cite{Alver:2007wy}.}.  Both models are observed to
roughly reproduce the strength of non-flow correlations in p+p
collisions at large $\deta$. Due to large systematic uncertainties in
the p+p data, \hij\ simulations are used to model the long range
non-flow correlations in Au+Au collisions by assuming non-flow
correlations in data are some multiplicative factor, $m$, times the
non-flow in \hij\ ($\delta_{\rm MC}(\eta_1,\eta_2)$) for
$\detagt$. This can be incorporated by modifying \eq{eq:v2sqrfit}:
\begin{multline}
  \vtwosqr - m\delta_{\rm MC}(\eta_1,\eta_2)= \\
  \vtwoetaf{1}\times \vtwoetaf{2} \: ; \: \detagt.
  \label{eq:v2sqrfitmod}
\end{multline}

\begin{figure}[t]
  \centering
  \includegraphics[width=0.47\textwidth]{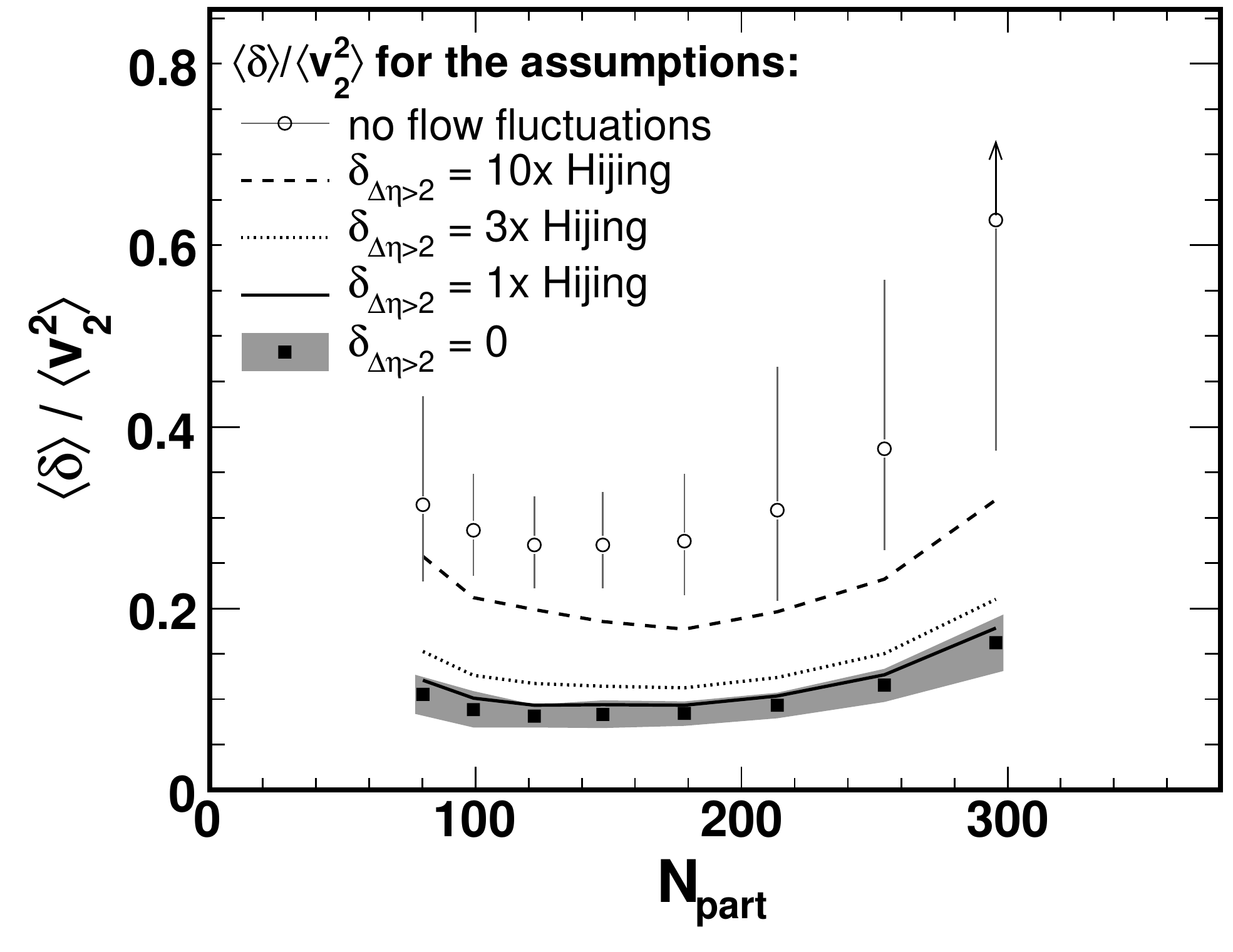}
  \caption{The non-flow ratio ($\mean{\delta}/\mean{v_2^2}$) in the \phob\
    Octagon detector acceptance as a function of number of
    participating nucleons (\Npart) in Au+Au collisions at $\snn
    =$~200~GeV. The black squares show the results with the assumption
    that non-flow correlations are negligible at $\detagt$. The shaded
    band shows the 90\% confidence systematic errors. The lines
    show different assumptions about non-flow at $\detagt$. The open
    circles with 90\% C.L. systematic errors, show the upper limit on
    $\mean{\delta}/\mean{v_2^2}$ obtained by assuming that the measured
    dynamic fluctuations in $\vtwo$ are due to non-flow alone.}
  \label{fig:deltaovervsnparta}
\end{figure}

The resulting non-flow ratio, $\mean{\delta}/\mean{v_2^2}$, found by
applying \mbox{Eqs.~\ref{eq:subtracttogetdelta}-\ref{eq:averagev22}}
with the modified $\vtwoetaf{}$ results, is plotted as a function of
centrality in \fig{fig:deltaovervsnparta} for different assumptions on
the magnitude of non-flow at $\detagt$. 
If non-flow correlations are assumed to be present only in $\detalt$
($m=0$), it is found that they account for approximately 10\% of the
observed $v_2^2$ signal averaged over $|\eta|<3$.
The results do not change significantly if the long range non-flow
correlations ($\delta_{\detagt}$) are taken to be the same as the
correlations in \hij\ ($m=1$ instead of $m=0$).

The upper limit on the non-flow ratio, also shown in
\fig{fig:deltaovervsnparta}, is drawn from the measurement of dynamic
$v_2$ fluctuations~\cite{Alver:2007qw} assuming that the observed
fluctuations are all due to non-flow correlations. The calculation of
this limit is described in Appendix~\ref{app:relate}.  This limit
corresponds to non-flow correlations in Au+Au collisions that are more 
than an order of magnitude higher than the expected correlations from
p+p collisions for $\detagt$ ($m>10$).

\section{Elliptic flow fluctuations}
\label{sect:flow}

An event-by-event measurement of the anisotropy in heavy ion collisions
yields fluctuations from three sources: statistical fluctuations due
to the finite number of particles observed, elliptic flow fluctuations
and non-flow correlations. We have previously measured the dynamic
fluctuations in $v_2$ by taking out the statistical fluctuations with
a study of the measurement response to the input $v_2$
signal~\cite{Alver:2007qw}. The new results on the magnitude of
non-flow correlations presented in the previous section can be used to
decouple the contributions of genuine elliptic flow fluctuations and
non-flow correlations to the measured dynamic fluctuations.

Let us denote the observed distribution of the event-by-event
anisotropy as $g(\vtwoobspaper)$, the distribution of the intrinsic elliptic
flow value as $f(\vtwo)$ and the expected distribution of $\vtwoobspaper$
for a fixed value of $\vtwo$ as $K(\vtwoobspaper,\vtwo)$.  We assume
$f(\vtwo)$ to be a Gaussian in the range $\vtwo>0$ with two
parameters, mean~($\mean{\vtwo}$) and standard
deviation~($\sigma$). The dynamic fluctuations in $\vtwo$, can be
calculated by unfolding the experimental measurement
$g^\expt(\vtwoobspaper)$ with a response function $K_n^\expt(\vtwoobspaper,\vtwo)$
which accounts for detector effects and statistical fluctuations:
\begin{equation}
  g^\expt(\vtwoobspaper) =   \int_{0}^{1} K_n^\expt(\vtwoobspaper,\vtwo)
  f_{\text{dyn}}(\vtwo) \der \vtwo.
\end{equation}
The calculation of intrinsic flow fluctuations
($f_{\text{flow}}(\vtwo)$) from measured dynamic fluctuations
($f_{\text{dyn}}(\vtwo)$) can be summarized by the following equation:
\begin{multline}
  \int_{0}^{1} K_{n}(\vtwoobspaper,\vtwo) f_{\text{dyn}}(\vtwo) \der \vtwo \\
  = \int_{0}^{1} K_{n,\delta}(\vtwoobspaper,\vtwo) f_{\text{flow}}(\vtwo)
  \der \vtwo,
  \label{eq:addinquadsum}
\end{multline}
where $K_{n}(\vtwoobspaper,\vtwo)$ and
$K_{n,\delta}(\vtwoobspaper,\vtwo)$ are the response functions for an
ideal detector with and without non-flow correlations respectively.
\Eq{eq:addinquadsum} gives the distribution of observed anisotropy for
an ideal detector $g(\vtwoobspaper)$, such that on the left hand side
the non-flow correlations are encoded in the dynamic $\vtwo$
fluctuations, and on the right hand side, they are accounted for in
the response function $K_{n,\delta}(\vtwoobspaper,\vtwo)$.
The response functions $K_{n}(\vtwoobspaper,\vtwo)$ and
$K_{n,\delta}(\vtwoobspaper,\vtwo)$ are given by a
Bessel-Gaussian distribution~\cite{Ollitrault:1992bk} defined as
\begin{multline}
  \BesselGaus{\vtwoobspaper}{\vtwo}{\sigma_s} \equiv \frac{\vtwoobspaper}{\sigma_s^2} \\
  \times \exp\left(-\frac{(\vtwoobspaper)^2+\vtwo^2}{2\sigma_s^2}\right)
  I_{0} \left(\frac{\vtwoobspaper \vtwo}{\sigma_s^2}\right),
 \label{eq:besselgaus}
\end{multline}
where $I_{0}$ is the modified Bessel function. The fluctuation term
$\sigma_s$ in the response function is a quadratic sum of statistical
fluctuations ($\sigmastat = 1/\sqrt{2n}$) due to finite number of
particles ($n$) observed in the detector and a contribution from
non-flow correlations ($\sigmadelta = \sqrt{\delta/2}$).

\begin{figure}[t]
  \includegraphics[width=0.47\textwidth]{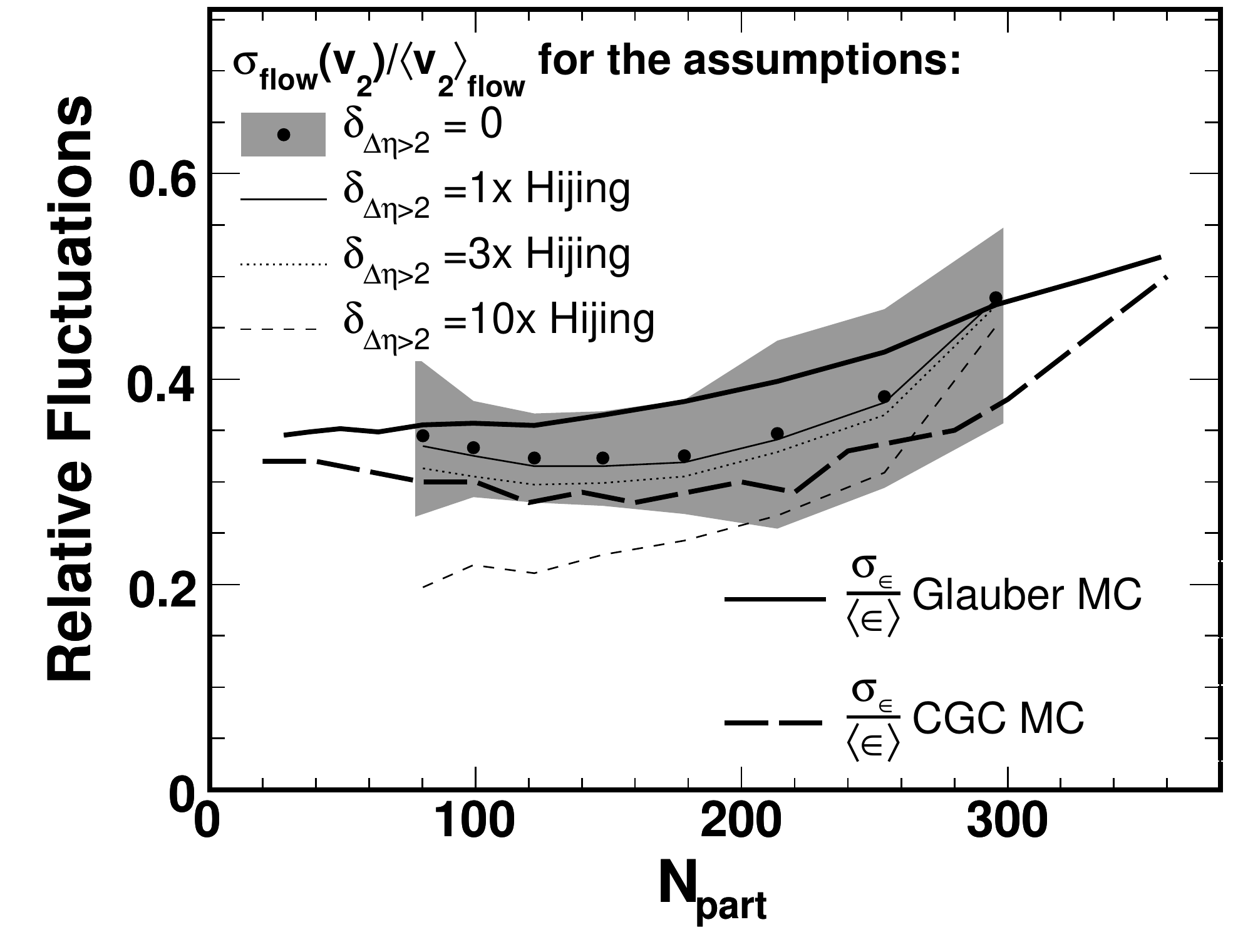}
  \caption{Relative elliptic flow fluctuations
    ($\sigmaflow/\meanvtwoflow$) as a function of number of
    participating nucleons (\Npart) in Au+Au collisions at $\snn
    =$~200~GeV.  The black circles show the results with the assumption
    that non-flow correlations are negligible at $\detagt$. The shaded
    band shows the 90\% confidence systematic errors. The thin lines
    show results for different assumptions on the magnitude of
    non-flow at $\detagt$.  The continuous and dashed thick lines show
    $\sigma(\epsilon_{part})/\langle\epsilon_{par t}\rangle$ values
    calculated in Glauber MC~\cite{Alver:2007qw} and
    CGC~\cite{Drescher:2007ax} models, respectively.}
\label{fig:flowecc}
\end{figure}

\Eq{eq:addinquadsum} cannot be simplified analytically. However, it
can be solved numerically to calculate relative elliptic
flow fluctuations ($\sigmaflow/\meanvtwoflow$) that correspond to the
measured dynamic $v_2$ fluctuations ($\sigmatot/\mean{\vtwo}$) and the
non-flow ratio ($\mean{\delta}/\mean{v_2^2}$) for different assumptions on non-flow
at $\detagt$. The details of the numerical calculation are given in
Appendix~\ref{app:relate}. It has been suggested that the relation
between these quantities can be approximated as
$\sigmatot^2=\sigmadelta^2+\sigmaflow^2$~\cite{Ollitrault:2009ie}.  We
have found that this approximation does not hold in the range of our
experimental results ($\sigmatot/\mean{\vtwo}>0.3$).

The systematic error in the magnitude of relative elliptic flow
fluctuations is obtained by propagating the errors in the measured
quantities $\sigmatot/\mean{\vtwo}$ and $\mean{\delta}/\mean{v_2^2}$ and by varying
the procedure to calculate $\sigmaflow/\meanvtwoflow$ from these
quantities. The errors from different sources are added in quadrature
to obtain the 90\% confidence interval. The error propagated from the
uncertainty in $\sigmatot/\mean{\vtwo}$ is the dominant contribution
to the uncertainty in $\sigmaflow/\meanvtwoflow$.

The relative fluctuations in the event-by-event elliptic flow,
corrected for contribution of non-flow correlations are presented in
\fig{fig:flowecc} as a function of the number of participating
nucleons, in Au+Au collisions at $\snn =$ 200~GeV for 6--45\% most
central events.  The elliptic flow fluctuations are found to be
roughly 30--40\% if the magnitude of non-flow correlations are assumed
to be small for $\detagt$.  The observed values of relative elliptic
flow fluctuations correspond to 87-97\% (79-95\%) of the previously
measured dynamic $v_2$ fluctuations~\cite{Alver:2007qw} if non-flow
correlations at $\detagt$ are assumed to be zero (three times the
magnitude in \hij).

Also shown in \fig{fig:flowecc} are relative fluctuations in the
participant eccentricity obtained from MC Glauber~\cite{Alver:2007qw}
and color glass condensate(CGC)~\cite{Drescher:2007ax} calculations.
The measured values of elliptic flow fluctuations are observed to be
consistent with both models over the centrality range under study if
the long range non-flow correlations are neglected.  The same
conclusion holds if the long range correlations are assumed to be
three times stronger than in p+p collisions, as modeled by \hij.

\section{Summary and conclusions}
\label{sect:conclude}
We have presented new data on the magnitude of non-flow
correlations and the event-by-event elliptic flow
fluctuations corrected for non-flow correlations in Au+Au collisions at
$\snn =$~200~GeV.  The measurement of non-flow
correlations is achieved by utilizing a new correlation analysis with
the assumption that non-flow correlations are of the order that is
observed in p+p collisions for long range correlations
($\detagt$).   The non-flow correlations averaged over the \phob\
Octagon acceptance ($-3 \! < \! \eta \! < \! 3$) are found to be
large, constituting approximately 10\% of the measured
$v_2^2$ signal.
Studying the dependence of expected azimuthal anisotropy fluctuations
due to non-flow correlations, it is found that the long range non-flow
correlations in Au+Au collisions would have to be more than an order of
magnitude stronger compared to the p+p data for non-flow correlations
to lead to the observed azimuthal anisotropy fluctuations with no
intrinsic elliptic flow fluctuations.
The method presented in this paper can be generally
applied in large acceptance detectors to study the contribution of
non-flow correlations to the flow signal measured with different
approaches.

The magnitude of event-by-event elliptic flow fluctuations were
calculated by subtracting the contribution of non-flow correlations to
the measured values of dynamic $v_2$ fluctuations. If the inclusive
long range non-flow
correlations in A+A collisions are assumed to be of the order of
magnitude that is observed in p+p collisions, the magnitude of
event-by-event elliptic flow fluctuations are found to be in agreement
with predicted fluctuations of the initial shape of the collision
region in both Glauber and Color Glass Condensate models.  Therefore
these results support conclusions from previous studies on the
importance of geometric fluctuations of the initial collision region
postulated to relate elliptic flow measurements in the Cu+Cu and Au+Au
systems~\cite{Alver:2006wh}.

%
%
%
%

This work was partially supported by U.S. DOE grants 
DE-AC02-98CH10886,
DE-FG02-93ER40802, 
DE-FG02-94ER40818,  
DE-FG02-94ER40865, 
DE-FG02-99ER41099, and
DE-AC02-06CH11357, by U.S. 
NSF grants 9603486, 
0072204,            
and 0245011,        
by Polish MNiSW grant N N202 282234 (2008-2010),
by NSC of Taiwan Contract NSC 89-2112-M-008-024, and
by Hungarian OTKA grant (F 049823).

\begin{figure}[t]
  \includegraphics[width=0.47\textwidth]{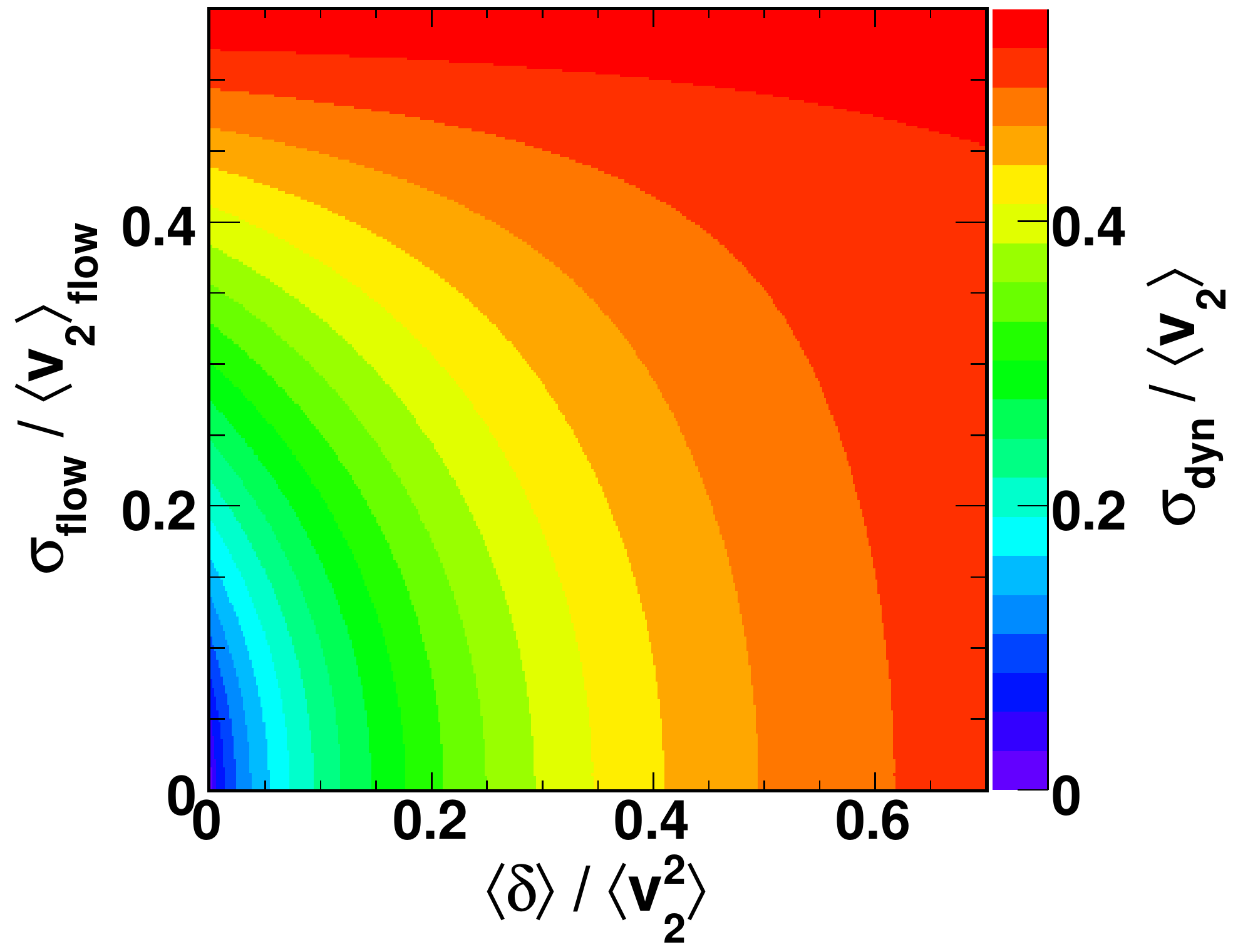}
  \caption{Dynamic $v_2$ fluctuations ($\sigmatot/\mean{\vtwo}$) as a
    function of elliptic flow fluctuations
    ($\sigmaflow/\meanvtwoflow$) and the non-flow ratio
    ($\mean{\delta}/\mean{v_2^2}$) for
    $\sigmastat/\meanvtwoflow=0.6$. The observed values of dynamic
    $v_2$ fluctuations are roughly given by
    $\sigmatot/\mean{\vtwo}\approx$40\%~\cite{Alver:2007qw}.}
  \label{fig:relations}
\end{figure}

\appendix
\section{Numerical calculations relating measured quantities to
  elliptic flow fluctuations}
\label{app:relate}

In this section, we describe the numerical calculations performed to
relate the measured values of dynamic $v_2$ fluctuations
($\sigmatot/\mean{\vtwo}$) and
non-flow ratio ($\mean{\delta}/\mean{v_2^2}$) to intrinsic elliptic flow
fluctuations ($\sigmaflow/\meanvtwoflow$).

We start by assuming the mean value of the elliptic flow distribution and the
magnitude of statistical fluctuations to be given as
$\meanvtwoflow=0.06$ and $\sigmastat=0.6\times \meanvtwoflow=0.036$ (see
\eq{eq:besselgaus}). Then, for
given values of $\sigmaflow/\meanvtwoflow$ and
$\mean{\delta}/\mean{v_2^2}$, the expected
distribution of the observed event-by-event anisotropy $\vtwoobspaper$
 can be calculated as
\begin{equation}
  g(\vtwoobspaper)=\int_{0}^{1} K_{n,\delta}(\vtwoobspaper,\vtwo) f_{\text{flow}}(\vtwo) \der \vtwo,
\end{equation}
where $f_{\text{flow}}(\vtwo)$ is a Gaussian in the range $\vtwo>0$
with mean and standard deviation values given by $\meanvtwoflow$ and
$\sigmaflow$, respectively, and $K_{n,\delta}(\vtwoobspaper,\vtwo)$ is given
by a Bessel-Gaussian (see~\eq{eq:besselgaus}),
\begin{equation}
  K_{n,\delta}(\vtwoobspaper,\vtwo)= \BesselGaus{\vtwoobspaper}{\vtwo}{\sigma_s}.
  \label{eq:kernelnodelta}
\end{equation}
The fluctuations encoded in the response function
$K_{n,\delta}(\vtwoobspaper,\vtwo)$ are given as
$\sigma_s^2=\sigmastat^2+\sigmadelta^2$, where $\sigmadelta$ can be
calculated from $\meanvtwoflow$, $\sigmaflow$ and
$\mean{\delta}/\mean{v_2^2}$:
\begin{eqnarray}
  2 \sigmadelta^2 &=& \mean{\delta}\\
  &=& \mean{\delta} \times \frac{\meanvtwoflow^2+\sigmaflow^2}{\mean{v_2^2}-\mean{\delta}}\\
  &=&  \frac{\mean{\delta}/\mean{v_2^2}}{1-\mean{\delta}/\mean{v_2^2}} \times (\meanvtwoflow^2+\sigmaflow^2).
\end{eqnarray}
In this derivation, it has been noted that the $\mean{\vtwo^2}$ defined in
\eq{eq:averagev22} includes contributions from flow fluctuations and
non-flow correlations.

Next, we calculate the dynamic fluctuations in the measured
$\vtwoobspaper$ distribution, $g(\vtwoobspaper)$, by using a response
function which incorporates only statistical fluctuations but not
non-flow correlations,
\begin{equation}
  g(\vtwoobspaper)=\int_{0}^{1} K_{n}(\vtwoobspaper,\vtwo) f_{\text{dyn}}(\vtwo) \der \vtwo.
  \label{eq:appdyn}
\end{equation}
Assuming the dynamic $\vtwo$ fluctuations are described by a Gaussian,
$f_{\text{dyn}}(\vtwo)$, in the range $\vtwo>0$ with mean and standard
deviation values given by $\mean{\vtwo}$ and $\sigmatot$, the value of 
 $\sigmatot/\mean{\vtwo}$ can be obtained by fitting \eq{eq:appdyn}. 

The resulting distribution of $\sigmatot/\mean{\vtwo}$ as a function
of $\sigmaflow/\meanvtwoflow$ and $\mean{\delta}/\mean{v_2^2}$ is shown
in \fig{fig:relations}. The value of $\sigmaflow/\meanvtwoflow$
corresponding to measured values of $\sigmatot/\mean{\vtwo}$ and
$\mean{\delta}/\mean{v_2^2}$ can be extracted from this
distribution. Furthermore, the values for $\sigmaflow/\meanvtwoflow=0$
can be used to set an upper limit on the magnitude of the non-flow ratio.

Since the related quantities are given as ratios, the value of
$\meanvtwoflow$ set at the beginning is arbitrary. It was observed that
$\sigmastat/\mean{v_2}$ is roughly given by $0.6$ for the dynamic
$v_2$ fluctuations measurement for all centrality bins in the
centrality range studied. The calculation was repeated for
values of $\sigmastat/\meanvtwoflow$=0.4 and 0.8. The differences in
results, which were found to be small, are incorporated in the
systematic errors.

\bibliographystyle{apsrev}
\bibliography{thesis}

\end{document}